\documentclass{mn2e}
\input{epsf}

\title[The slopes of density profiles in dSphs]
{Can we measure the slopes of density profiles in dwarf spheroidal galaxies?}

\author[K. Kowalczyk et al.]
    {Klaudia Kowalczyk,$^{1}$ Ewa L. {\L}okas,$^{2}$
    Stelios Kazantzidis$^{3}$ and Lucio Mayer$^{4}$
    \\
    \\
    $^1$Warsaw University Observatory, Al. Ujazdowskie 4, 00-478 Warsaw, Poland\\
    $^2$Nicolaus Copernicus Astronomical Center, Bartycka 18, 00-716 Warsaw, Poland  \\
    $^3$Center for Cosmology and Astro-Particle Physics; and Department of Physics; and Department of Astronomy,
    The Ohio State\\ University, Columbus, OH 43210, USA\\
    $^4$Institute for Theoretical Physics, University of Z\"urich, CH-8057 Z\"urich, Switzerland
    }

\begin{document}

\maketitle

\begin{abstract}
Using collisionless $N$-body simulations of dwarf galaxies orbiting the Milky Way we construct realistic models
of dwarf spheroidal (dSph) galaxies of the Local Group. The dwarfs are initially composed of stellar disks embedded in
dark matter haloes with different inner density slopes and are placed on an eccentric orbit typical for
Milky Way subhaloes. After a few Gyr of evolution the stellar component is triaxial as a result of bar instability
induced by tidal forces. Observing the simulated dwarfs along the three principal axes of the stellar component
we create mock data sets and determine the corresponding half-light radii and line-of-sight velocity dispersions.
Using the estimator proposed by Wolf et al. we calculate the masses within half-light radii. The masses obtained in this
way are over(under)estimated by up to a factor of two when the line of sight is along the longest (shortest) axis of
the stellar component. We then divide the initial stellar distribution into an inner and outer population and trace their
evolution in time. The two populations, although strongly affected by tidal forces, retain different density profiles
even after a few Gyr of evolution. We measure the half-light radii and velocity dispersions of the stars in the two
populations along different lines of sight and use them to estimate the slope of the mass distribution in the dwarf
galaxies following the method recently proposed by Walker \& Pe\~narrubia. The inferred slopes are systematically
over- or underestimated, depending on the line of sight. In particular, when the dwarf is seen along the longest axis
of the stellar component, a significantly shallower density profile is inferred than the real one measured from the
simulations. Given that most dSph galaxies in the Local Group are non-spherical in appearance and their orientation
with respect to our line of sight is unknown, but most probably random, the method can be reliably applied only to
a large sample of dwarfs when these systematic errors are expected to be diminished.
\end{abstract}


\section{Introduction}

One of the key predictions of theories of structure formation based on cold dark matter is that the dark matter
haloes should possess cuspy density profiles in the centre of bound structures such as galaxies and galaxy clusters
(Navarro, Frenk \& White 1997, hereafter NFW).
The prediction, obtained via the use of $N$-body simulations of gravitational instability of dark matter only, relied
on the assumption that baryons play relatively minor role in shaping the final density structure of bound objects.
Comparison of this prediction to observations of low surface brightness galaxies resulted in strong disagreement, as
modelling of rotation curves of such objects consistently pointed towards preference for cored, rather than cuspy
dark matter density profiles required by the data (e.g. de Blok, McGaugh \& Rubin 2001). The tension came to be known
as the cusp/core problem.

A number of solutions to the problem has been proposed over the years including the possibility that the
difficulties associated with modelling of the data prevent us from detecting the cusps which really are there and
the proposals to modify the properties of dark matter postulating that it may be self-interacting
(Spergel \& Steinhardt 2000; Vogelsberger, Zavala \& Loeb 2012). The most
effective approach, however, turned out to be the one questioning the aforementioned assumption of baryon physics
playing little or no role in shaping dark matter haloes. This approach has recently led to more detailed modelling
of star-formation related processes using $N$-body and hydro simulations in the cosmological context
and succeeded in producing disky dwarf galaxies with shallower inner dark matter density
slopes (Governato et al. 2010). The success of this work relied on including all the crucial baryonic processes like
gas cooling, star formation, cosmic UV background heating and gas outflows driven by supernovae but the critical
improvement involved resolving individual star-forming regions and allowing only high density gas clouds
to form stars. The predictions of this theoretical work were successfully compared to the properties of late-type
dwarf galaxies from the THINGS survey (Oh et al. 2011).

The question remains, if similar agreement can be reached for early-type galaxies, in particular dwarf spheroidals
(dSph) available for observation in the immediate vicinity of the Milky Way. In systems supported by random rather
than rotational motions, an additional complication in the modelling occurs due to the fact that such systems may
be characterized by a variety of velocity anisotropy profiles a priori unknown which leads to the degeneracy between
the model parameters. The inferences concerning dark matter distribution in dSph galaxies are usually made based on
solving the lowest-order Jeans equation that relates the mass distribution to observables such as the density
distribution of the stars and their velocity dispersion profile. Solving for the underlying mass distribution requires
however strong assumptions concerning the anisotropy of stellar orbits.

This degeneracy can be partially lifted by
including higher-order Jeans equations and modelling also the fourth velocity moment which is sensitive mainly to
velocity anisotropy ({\L}okas 2002). This strategy works if the dark matter profile is parametrized only
by the virial (or total) mass and concentration (or equivalently scale-length) and has been successfully applied to
the Draco dSph galaxy ({\L}okas, Mamon \& Prada 2005). However, if the inner dark matter slope is additionally
considered as a free parameter, another degeneracy occurs, this time between the inner slope and concentration, and
equally good fits to the velocity moments can be obtained for cuspy and cored dark matter profiles ({\L}okas \&
Mamon 2003; S\'anchez-Conde et al. 2007).

With the large kinematic samples currently available for some of the classic dSph galaxies of the Local Group it has
recently become possible to model their dark matter distribution using different approaches based on orbit superposition
and/or working with the distribution function itself and modelling of discrete data rather than velocity moments
which require binning of the data and thus lead to loss of information (Chanam\'e, Kleyna \& van der Marel 2008;
Jardel \& Gebhardt 2012; Breddels et al. 2012). The conclusions from this approach are far from consensus and
seem to depend on the object studied and the details of the method used. For example Breddels et al. cannot firmly
distinguish between a core and a cusp in the Sculptor dSph while Jardel \& Gebhardt claim to find preference for a
core-like dark matter distribution in Fornax. The subject thus requires further study and different approaches,
one promising example being the use of the motions of globular clusters in dSph galaxies (see Goerdt et al. 2006
and Cole et al. 2012 for an application of this method to the Fornax dwarf).

An interesting and particularly simple method to measure the slope of density profile in dSph galaxies has been
recently proposed by Walker \& Pe\~narrubia (2011, hereafter WP11). The method relies on the use of separate
stellar populations identified in a dSph by their different metallicity and simple mass estimators measuring the mass
within half-light radii of each population from these radii and their respective velocity dispersions. Since
both populations are expected to be in equilibrium in the global gravitational potential of the dwarf, the two
mass measurements at two different scales lead to a direct measurement of the mass slope which can be translated to
a constraint on the inner density slope. The application of this method to the Fornax and Sculptor dSph galaxies led
WP11 to the conclusion that both possess dark matter density profiles significantly shallower than the cuspy
profiles predicted by NFW.

In this work we test the method of WP11 using realistic models of dSph galaxies formed in $N$-body simulations of
tidal stirring of disky dwarfs embedded in dark matter haloes of different inner slopes in the gravitational field
of the Milky Way. The tidal stirring scenario
for the formation of dSph galaxies, originally proposed by Mayer et al. (2001), has been demonstrated to reproduce
well the basic properties of the population of dSph galaxies in the Local Group (Mayer et al. 2007;
Klimentowski et al. 2009a;
Kazantzidis et al. 2011; {\L}okas, Kazantzidis \& Mayer 2011). Although WP11 tested their
approach using $N$-body realizations of dwarf galaxies, they only used spherical models and concluded that any
systematic errors can lead to the underestimation of the mass slope and thus dwarfs cannot appear more core-like than
they really are.

However, dSph galaxies of the Local Group are known to be non-spherical, with an average ellipticity
of 0.3 (Mateo 1998; {\L}okas et al. 2011, 2012a). We expect this fact to have important consequences for the mass and
mass slope estimates. As already discussed by {\L}okas et al. (2010a), when the dwarfs are observed along the longest
(shortest) axis of the stellar component, their mass is significantly over(under)estimated even if modelling of
the velocity anisotropy is included in the analysis. The non-sphericity may affect simple mass estimators used in
the method of WP11 even more and thus bias the inferred mass slope estimates. In the simulations of tidal stirring
of dwarf galaxies
we use here non-spherical, triaxial stellar components form naturally as a result of bar instability induced by tidal
forces from the Milky Way and thus such models are suitable for testing the effects of non-sphericity on mass and
mass slope estimates.

The paper is organized as follows. In section 2 we describe the simulations used in this work and characterize
the main properties of the simulated dwarfs used for the further analysis. Section 3 is devoted to the problem
of mass estimation; we present the observables used, estimate the masses within half-light radii and compare them
to real masses measured directly from the simulations. In section 4 we describe the evolution of properties of two
stellar populations selected from the stellar component of the dwarfs and in section 5 we use them to estimate the
slope of the density profile in the simulated dwarfs. The discussion of the results follows in section 6.

\section{The simulated dwarfs}

In this work we used a subset of simulations described in more detail in {\L}okas, Kazantzidis \& Mayer (2012b).
The dwarf galaxy models consisted of
exponential stellar disks embedded in spherical dark matter haloes with density profiles of the form
\begin{equation}    \label{densityprofiles}
    \rho(r) = \frac{\rho_{\rm char}}{(r/r_{\rm
    s})^\alpha \,(1+r/r_{\rm s})^{3-\alpha}}.
\end{equation}
Such profiles are characterized by an asymptotic inner slope $r^{-\alpha}$, which we allow to vary,
and an outer slope $r^{-3}$ which is kept fixed. The detailed properties of density profiles given by
formula (\ref{densityprofiles}) are discussed in
{\L}okas (2002) and {\L}okas \& Mamon (2003) who applied them in dynamical modelling of dSph galaxies
and the Coma cluster. All our
dwarf galaxies have the same virial mass of $M_{\rm vir} = 10^{9} M_{\odot}$
and concentration parameter $c=r_{\rm vir}/r_{\rm s}=20$, but different $\alpha$,
which requires slightly different values of the characteristic density
$\rho_{\rm char}$ in equation (\ref{densityprofiles}).
We consider $\alpha=1$ (which corresponds to the
NFW profile), and two shallower inner slopes, a mild cusp with
$\alpha=0.6$ and an almost cored profile with $\alpha=0.2$.

The dark matter haloes were populated with stellar disks of mass equal to $m_{\rm d} = 0.02 M_{\rm vir}$. The disk scale
lengths were $R_{\rm d} = 0.41$ kpc in the radial direction (corresponding to a dimensionless
spin parameter of $\lambda=0.04$, following Mo, Mao \& White 1998), and $z_{\rm d} = 0.2 R_{\rm d} $
in the vertical direction. Each numerical realization of the dwarf galaxy contained $N_{\rm h} = 10^6$ dark matter and $N_{\rm d} =
5 \times 10^5$ disk particles and the adopted gravitational softenings were
$\epsilon_{\rm h}=60$~pc and $\epsilon_{\rm d}=20$~pc, respectively.

\begin{figure}
\begin{center}
    \leavevmode
    \epsfxsize=7.5cm
    \epsfbox[60 170 525 635]{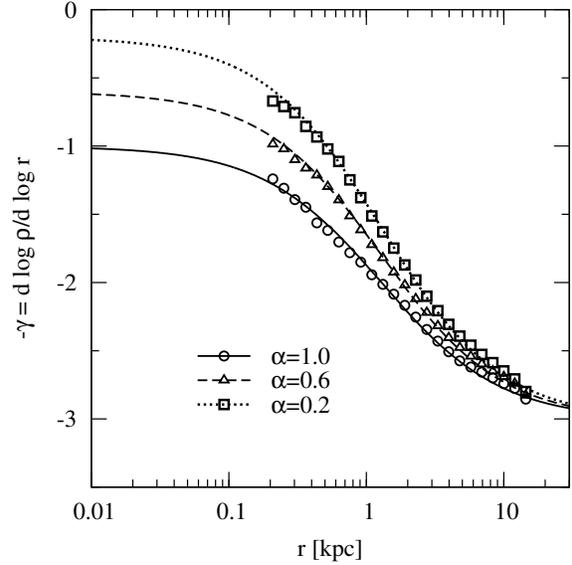}
\end{center}
\caption{The slope of the dark matter density profile of the haloes of the initial dwarf models
as a function of radius. Symbols show the values measured from the $N$-body realizations by fitting a power-law, while
the lines correspond to the values calculated from the analytic formula of equation (\ref{densityprofiles}).}
\label{gamma}
\end{figure}

It is worth emphasizing that the $\alpha$ denotes the {\em limiting\/} inner slope and thus differs from
the real slope one actually measures at any radius $r>0$. To illustrate this, in Figure~\ref{gamma} we plot
the dark matter density slope $-\gamma = {\rm d} \log \rho/{\rm d} \log r$ as a function of radius. The solid, dashed
and dotted lines show respectively the slope derived directly from formula (\ref{densityprofiles}). The symbols,
on the other hand, show the corresponding values of the slope measured from the numerical realizations of the
haloes by fitting a straight line to data points of $\log \rho(\log r)$. The latter measurements extend down to
radii of the order of 3 dark matter softening scales, i.e. about 0.2 kpc. The two measurements agree except for small
variations inherent in the numerical realizations.

At 0.2 kpc, and any larger radius, the actual slope is significantly lower than the limiting value $-\alpha$.
In particular, at $r=0.5$ kpc the dark matter density slopes of our dwarfs are initially in the range $(-1.6, -1)$
(see Figure~\ref{gamma}) and thus similar to those predicted
by the recent simulations of Governato et al. (2012) for dwarfs of stellar masses of the order of $2 \times 10^7$
M$_{\odot}$ at redshift $z=0$. Given that the inner slopes of dark matter haloes
are not significantly altered after redshift of about $z=1$
(Governato et al. 2010) we thus conclude that our assumed dark matter profiles are consistent with the latest findings
concerning the properties of dwarfs formed in cosmological simulations where star formation processes are
sufficiently resolved. Our progenitor dwarfs can therefore be considered as idealized versions of such dwarfs
which we assume were accreted by a Milky Way-like host at about $z=1-2$.

The dwarf galaxies were placed on an orbit (at apocentre) around
a primary galaxy with the present-day
structural properties of the Milky Way (Widrow \& Dubinski 2005; Kazantzidis et al. 2011) and their evolution was
followed for 10 Gyr using the $N$-body code PKDGRAV (Stadel 2001). The dwarf galaxy disk was oriented so that
its internal angular momentum was inclined to the orbital angular momentum by $i=45^{\circ}$.
Out of five orbits of different size and eccentricity considered in {\L}okas et al. (2012b) we choose only one,
namely R1, with orbital apocentre $r_{\rm apo}=125$ kpc and pericentre $r_{\rm peri}=25$ kpc.
This choice was motivated by the fact that only for this orbit
for all values of $\alpha$ at some stage of the evolution a dSph galaxy is formed and it survives for long enough
to provide sufficient number of models as input for the present analysis. We also note that this is a
typical orbit of a satellite accreted by a Milky Way-like galaxy at redshift $z=1-2$ (Diemand
et al. 2007; Klimentowski et al. 2010).

\begin{figure*}
\begin{center}
    \leavevmode
    \epsfxsize=16.cm
    \epsfbox[58 525 580 820]{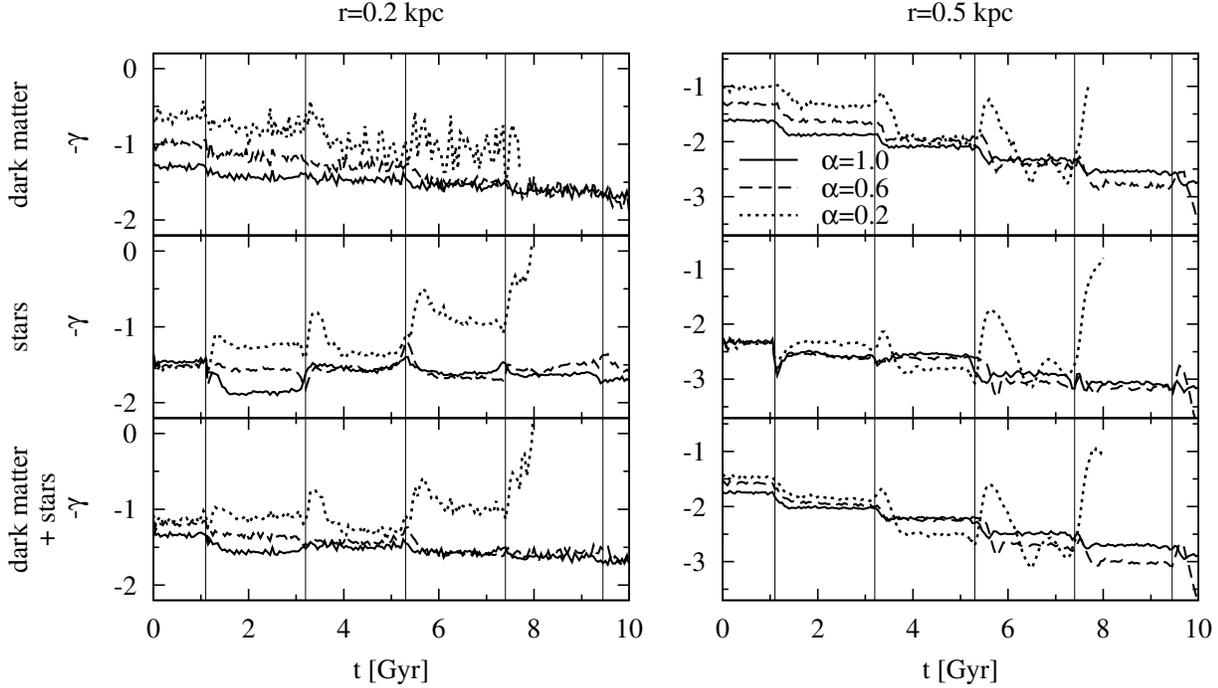}
\end{center}
\caption{The evolution of the slope of the density profile of different components in time
measured at the radius $r=0.2$ (left
panels) and $r=0.5$ kpc (right panels) from the centre of the dwarf. The rows from top to bottom
give results for dark matter, stars and the two components combined.
The solid, dashed and dotted lines show slopes measured for dwarfs with
haloes of different initial inner slope $\alpha = 1, 0.6$ and 0.2, respectively. Thin vertical lines indicate
pericentre passages.}
\label{evolution}
\end{figure*}

Let us first consider the evolution of the slope of the density profile of different components in time,
as the dwarfs orbit the Milky Way, since this is the primary focus of this work.
Figure~\ref{evolution} shows how the local slope of density profile of dark matter, stars and
the two components combined (from the top to the bottom row)
changes due to tidal effects. In the two columns we plot the slopes measured at two scales, $r=0.2$ (left panels)
and $r=0.5$ kpc (right panels) which bracket the scales where the measurements from the kinematic data can be
performed since these are of the order of the half-light radii of our simulated dwarf galaxies at later stages
of evolution. Note that we do not consider here the limiting slope of the dark matter density profile at
$r \rightarrow 0$, as was done e.g. by Kazantzidis et al. (2004), but we measure the slope at a fixed, non-zero
radius.

\begin{figure}
\begin{center}
    \leavevmode
    \epsfxsize=8.3cm
    \epsfbox[75 70 500 820]{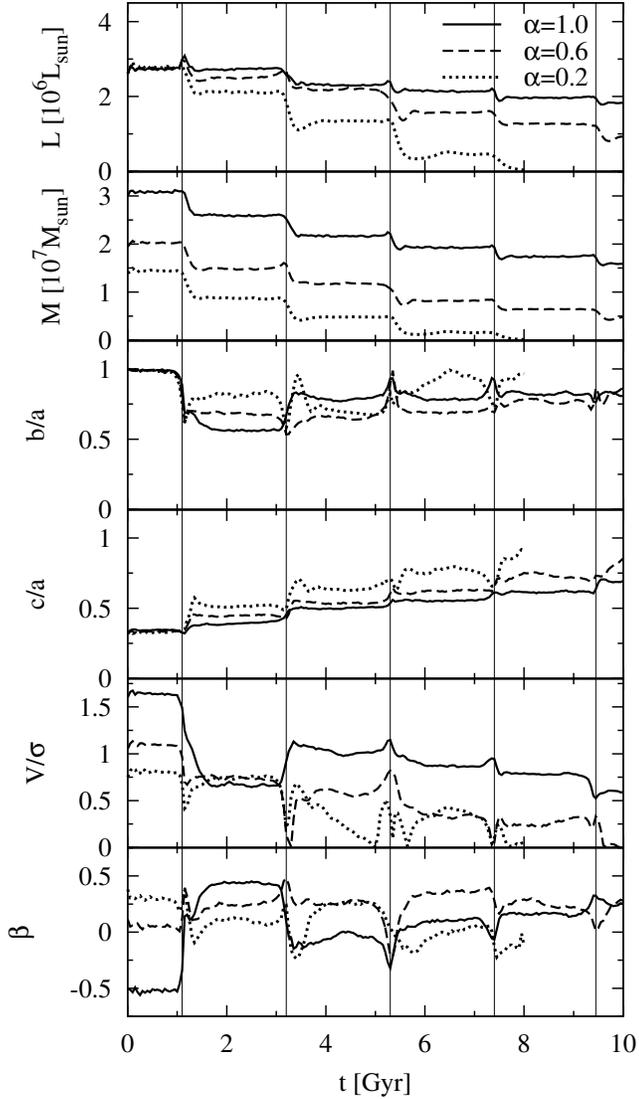}
\end{center}
\caption{The evolution of different properties of the dwarf galaxy in time. The panels from top to bottom plot
the luminosity, mass (of stars and dark matter), axis ratios, the ratio of the
rotation velocity to the velocity dispersion $V/\sigma$ and the anisotropy parameter $\beta$ of the stellar component.
All quantities
were measured within radius $r < 0.5$ kpc. In each panel the solid, dashed and dotted lines
show the results for dwarfs embedded in dark matter haloes with different initial inner slope $\alpha = 1, 0.6$
and 0.2, respectively. Thin vertical lines indicate pericentre passages.}
\label{transformation}
\end{figure}

The results demonstrate that in the case of dark matter
the slopes evolve towards stepper ones with time which reflects the overall steepening of the profiles due to
tidal stripping. In addition, the hierarchy of values of the slopes for different $\alpha$ is preserved, i.e. the local
slopes are always higher for $\alpha = 0.2$ than for $\alpha = 1$, except for the later stages when the slope
at $r=0.5$ kpc for the $\alpha = 0.2$ case varies strongly with time. This occurs when this dwarf galaxy starts
to dissolve before it is completely destroyed at about $t = 8$ Gyr, which is why we do not show results beyond
that time for this case in Figure~\ref{evolution} and any of the following plots.

In the case of the stellar component
the interpretation of the evolution of the slope is less straightforward because of the bar formation after the first
pericentre passage. However, at $r=0.5$ kpc the stellar density profile also steepens due to mass loss. In the inner
region, at $r=0.2$ kpc, the slope remains roughly constant in time and in the case of $\alpha=0.2$ even becomes shallower.
In general, at a given scale the dark matter density profile is shallower than the stellar one in the first stages
of evolution, reflecting the initial conditions, while the two profiles follow each other at the later stages.

The evolution of other properties of the dwarf galaxies with different $\alpha$ on our chosen orbit
is illustrated in Figure~\ref{transformation} (see Kazantzidis, {\L}okas \& Mayer 2013 for a more thorough
discussion for different orbits).
All quantities shown in the Figure were measured for particles inside $r<0.5$ kpc,
which again is motivated by the scale of maximum half-light radii at which the following analysis will be
performed. The two upper panels of the Figure illustrate the mass loss the dwarfs suffer, both in terms of
stellar content (first panel) and the total mass of stars and dark matter within $r<0.5$ kpc (second panel). To convert
the mass of stars to the luminosity we assumed the stellar mass-to-light ratio of 2.5 M$_{\odot}$/L$_{\odot}$
as is appropriate for an old, single-starburst stellar population typically hosted by dSph galaxies
(Bruzual \& Charlot 2003). As is immediately seen from the two panels, the mass is more effectively stripped
for lower $\alpha$ as the stars and dark matter are more weakly bound in these haloes (see the discussion in {\L}okas
et al. 2012b). In the case of $\alpha=0.2$ after the fourth pericentre, at about $t=8$ Gyr, the dwarf dissolves
completely and forms an elongated stream of uniform stellar density. Both $\alpha=0.6$ and $\alpha=1$ dwarfs
survive until the end of the simulation with final properties akin to the classical dSph galaxies of the Local
Group. In particular, none of those would qualify as an ultra-faint dwarf as is the case for tighter orbits
and $\alpha=0.6$ (see {\L}okas et al. 2012b).

The evolution of the shape of the stellar component of the dwarfs is illustrated in the third and fourth panel of
Figure~\ref{transformation} where we plot the axis ratio $b/a$ and $c/a$ where $a,b$ and $c$ are the longest,
intermediate and shortest axis determined using the inertia tensor for stars within $r<0.5$ kpc. All dwarfs
experience qualitatively the same evolution of the shape: after the first pericentre passage a triaxial shape
(or a bar) is formed which becomes more and more spherical in time. The transition towards the spherical shape
happens marginally faster for $\alpha=0.2$, especially in terms of $c/a$. The last two panels of
Figure~\ref{transformation} illustrate the evolution of the kinematics.
The ratio $V/\sigma$ measures the
amount of ordered versus random motion: $V$ is the mean rotation velocity around the shortest axis, while $\sigma$
is the 1D velocity dispersion obtained by averaging the dispersion measured along three spherical coordinates.
The anisotropy parameter $\beta$ is calculated in the standard way and includes rotation in the second velocity
moment around the shortest axis.

The evolution of the shape and kinematics shown in Figure~\ref{transformation} illustrates the transition from
the initial disks to dSph galaxies. It is customary to assume that a dSph galaxy is formed when the amount of
rotation is sufficiently diminished below some threshold (typically $V/\sigma < 1$),
and the shape is sufficiently close to spherical (for example $b/a > 0.5$ and $c/a > 0.5$).
Due to different initial properties our dwarfs have different $V/\sigma$ values at the beginning when
measured at a fixed radius. In particular at $r=0.5$ kpc the $\alpha=0.2$ has the lowest $V/\sigma < 1$
in spite of having a disk because the rotation curve rises slowly with radius as there is less mass in the centre
than for other $\alpha$ values. We thus modify the criterion and assume that the dSph galaxy is formed when
$V/\sigma$ drops below half of the initial value. For the shape criterion we adopt as usual the threshold of both
$b/a$ and $c/a$ greater than 0.5.

\section{Estimating masses}

Using these criteria for the formation of dSph galaxies
we selected 50 outputs (in the time range $t=7.5-10$ Gyr) for $\alpha=1$, 91 outputs
($t=5.5-10$ Gyr) for $\alpha=0.6$ and 22 outputs ($t=4.1-5.1$ Gyr) for $\alpha=0.2$ from the total of 201
outputs saved for each simulation. In the case of $\alpha=0.2$ there are actually more outputs satisfying the
criteria, but since after the third pericentre passage the dwarf is strongly perturbed and may departure from
equilibrium (as suggested by the strongly varying slope of the density profile in Figure~\ref{evolution})
we restrict the sample to earlier outputs as specified above.

For each selected output the dwarf was rotated so that the $x$ axis was oriented along the major, the $y$ axis
along the intermediate, and the $z$ axis along the shortest axis of the stellar component. The dwarf galaxy was
then ``observed", as a distant observer at infinity would do, along these three axes and mock data sets
including the stellar positions and velocities were created for each line of sight. The stellar positions were
binned in projected radius $R$ to measure the number density profile. We used bins equally spaced in $\log R$
and to each such profile we fitted the projected Plummer distribution
\begin{equation}	\label{plummer}
	\Sigma (R) = \frac{R_{\rm h}^2 N}{\pi (R^2 + R_{\rm h}^2)^2},
\end{equation}
adjusting the projected half-light radius $R_{\rm h}$ and normalization $N$. Examples of the measured and fitted
profiles (for one output for each $\alpha$) are shown in Figure~\ref{profiles} as triangles and solid lines,
respectively. The values of $R_{\rm h}$ for all selected outputs for different $\alpha$ are plotted as a
function of time in Figure~\ref{populations} as solid lines.

We then measured the line-of-sight velocity dispersion $\sigma_{\rm los}$ within $R_{\rm h}$
applying in each case a $3 \sigma$ clipping procedure to remove interloper stars until convergence was reached
and no more stars were removed from the sample. Note that within $R_{\rm h}$ the contamination from the tidal
tails in the immediate vicinity of the dwarf is expected to be very low (Klimentowski et al. 2007, 2009b) and therefore
the procedure removes mostly stars
with velocities very different from the mean velocity of the dwarf, belonging to tidal debris stripped much earlier.
The values of $\sigma_{\rm los}$ measured in this way are plotted as solid lines as a function of time
in Figure~\ref{sigma_populations} for different $\alpha$ and different lines of sight.

\begin{figure*}
\begin{center}
    \leavevmode
    \epsfxsize=17cm
    \epsfbox[75 405 560 590]{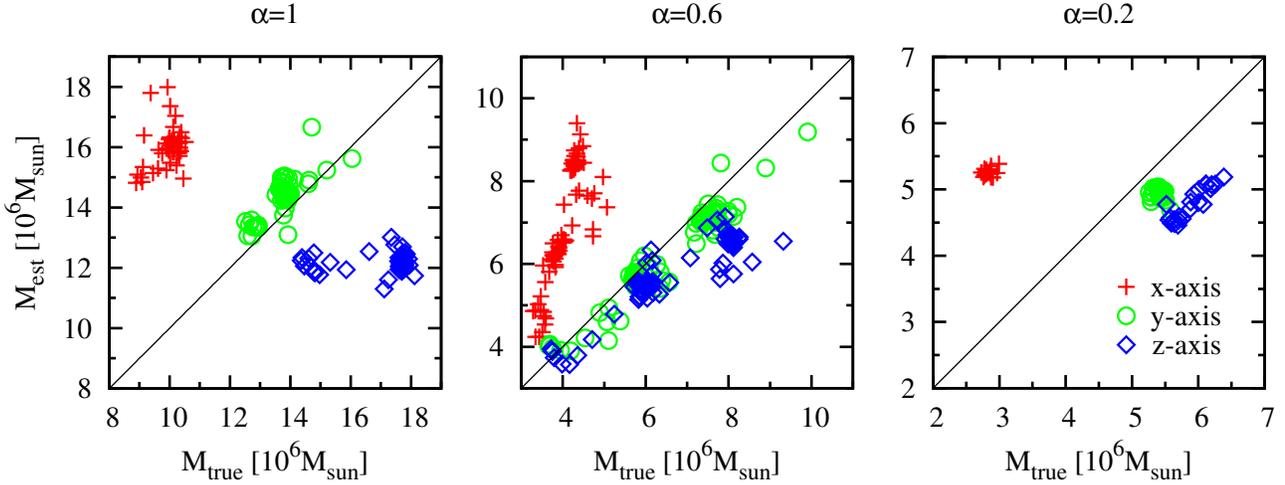}
\end{center}
\caption{Comparison between the masses $M_{\rm est}$ obtained using estimator (\ref{wolf})
and by direct measurement in the simulation $M_{\rm true}$.
The three panels from the left to the right show results for $\alpha = 1, 0.6$ and 0.2, respectively, and the diagonal
line in each panel corresponds to the equality of the masses.
Symbols of different shape and colour (red crosses,  green circles and blue squares)
indicate respectively observations performed along different
lines of sight: $x$ (longest axis of the stellar component), $y$ (intermediate axis) and $z$ (shortest axis).
For each line of sight the half-light radius $R_{\rm h}$ and velocity dispersion $\sigma_{\rm los}$ within
$R_{\rm h}$ were calculated using all stars in the appropriate region.}
\label{masses}
\end{figure*}

\begin{figure}
\begin{center}
    \leavevmode
    \epsfxsize=8.3cm
    \epsfbox[70 250 500 780]{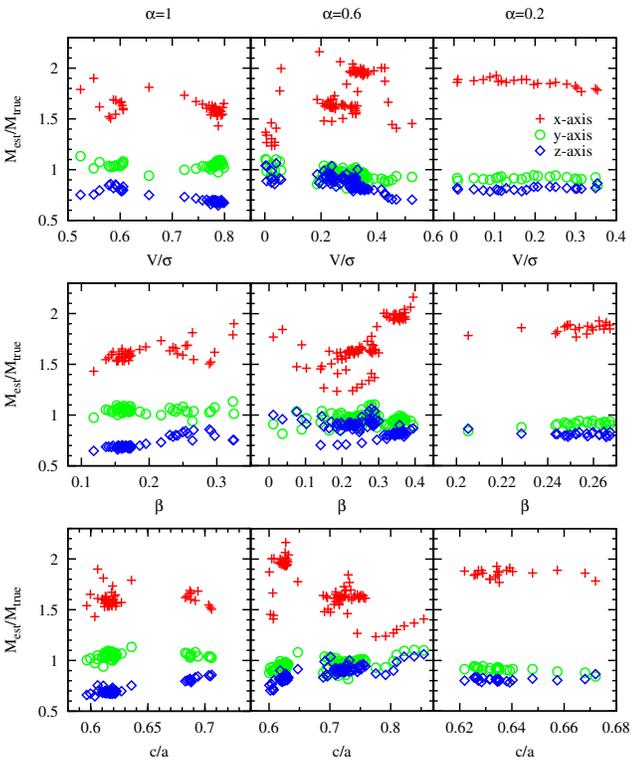}
\end{center}
\caption{The ratio of the estimated to true mass for all stars as a function of the amount of rotation $V/\sigma$
(upper row), the anisotropy parameter $\beta$ (middle row) and shape in terms of the ratio of the shortest
to longest axis of the stellar component $c/a$ (lower row). The columns from the left to the right show results
for $\alpha = 1, 0.6$ and 0.2, respectively. Symbols of different shape and colour indicate observations performed
along different lines of sight: $x$ (longest axis of the stellar component), $y$ (intermediate axis) and $z$
(shortest axis).}
\label{massratios}
\end{figure}

Using these measurements we estimated the masses of the dwarf galaxies in each output using the formula
proposed by Wolf et al. (2010)
\begin{equation}	\label{wolf}
	M_{\rm est} (r_3) = 3 \ G^{-1} \sigma_{\rm los}^2 r_3 = 3.7 \ G^{-1} \sigma_{\rm los}^2 R_{\rm h},
\end{equation}
where $r_3$ is the radius of the order of the 3D half-light radius found to give results least dependent on
anisotropy. For the Plummer profile we use here $r_3$ is related to the 3D half-light radius $r_{\rm h}$
and the 2D projected half-light radius $R_{\rm h}$
by $r_3/r_{\rm h}=0.94$ and $r_{\rm h}/R_{\rm h} = 1.305$, so $r_3 = f R_{\rm h}$ where $f=1.23$ (see the Appendix of
Wolf et al. 2010). The values of the mass estimated in this way were then compared to the real masses contained
within $f R_{\rm h}$ in the simulated dwarfs; we will refer to those masses as $M_{\rm true}$.

The values of $M_{\rm est}$ versus $M_{\rm true}$ are plotted in Figure~\ref{masses}. The three panels show the
results for all outputs selected for $\alpha = 1, 0.6$ and 0.2. We clearly see that while for the observations
along the intermediate $y$ axis of the stellar component (green circles) the estimated masses are quite close to
the corresponding true values, for observations along the longest axis $x$ the masses calculated from formula
(\ref{wolf}) are
overestimated and for observations along the shortest axis $z$ they are underestimated. Only for the observation
along the intermediate axis $y$ the agreement between $M_{\rm est}$ and $M_{\rm true}$ is good. The mean and dispersion
values of $M_{\rm est}/M_{\rm true}$ are listed in Table~\ref{estimated} in the first row for each $\alpha$.
The average numbers show that the masses can be systematically overestimated by up to almost a factor of 2 for
the observation along the $x$ axis and underestimated by up to 30 percent for the observation along $z$.

The reason for this systematic bias can be traced to the dependence of the estimates of $R_{\rm h}$ and
$\sigma_{\rm los}$ on the line of sight. As shown in Figures~\ref{populations} and \ref{sigma_populations},
when going from the line of sight along the $x$ axis to the $y$ and $z$ axis, the estimated half-light radius
increases, while the velocity dispersion decreases. As the mass estimator (\ref{wolf}) depends on the
combination $\sigma_{\rm los}^2 R_{\rm h}$, the direction of the bias is not a priori obvious. It turns out
however, that the effect is dominated by the input from the velocity dispersion since it is largest for the
line of sight along the longest ($x$) axis and the mass is overestimated in this case.

\begin{figure*}
\begin{center}
    \leavevmode
    \epsfxsize=16cm
    \epsfbox[70 415 575 620]{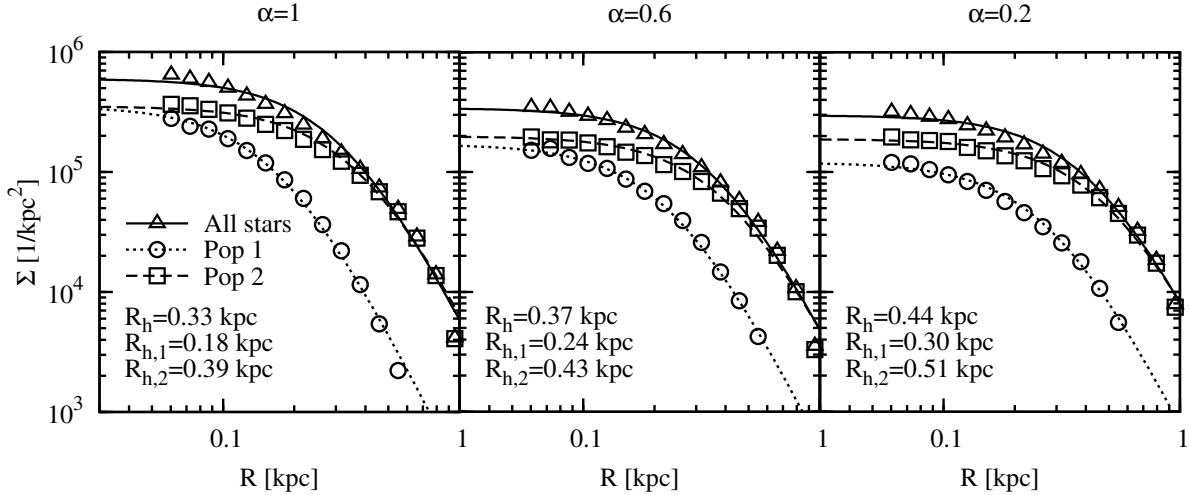}
\end{center}
\caption{Examples of measured surface density profiles of stars (symbols) and the corresponding fits of the
projected Plummer distribution (\ref{plummer}). The three panels from the left to the right show results for a single
simulation output in runs with $\alpha = 1, 0.6$ and 0.2, respectively. In each case the measurements were done
along the $y$ axis of the stellar component and at times $t=9.7$, 8.95 and 4.9 Gyr, for $\alpha = 1, 0.6$ and 0.2,
respectively. The fitted values of the half-light radii
for the two populations and for all stars are listed in the left corner of each panel.}
\label{profiles}
\end{figure*}

In Figure~\ref{massratios} we explore how the bias depends on three properties of the dwarfs, namely
the amount of rotation $V/\sigma$, the anisotropy parameter $\beta$ and shape in terms of the ratio of the shortest
to longest axis of the stellar component $c/a$. While there is no clear dependence on $V/\sigma$ and anisotropy,
there is a trend visible (for $\alpha =1$ and 0.6) with respect to shape such that the estimated masses
(especially for the most problematic line of sight along $x$) tend to be less biased for more spherical stellar
components.

\section{Stellar populations}

The method of estimating the slope of mass distribution in dSph galaxies proposed by WP11
relies on the use of two stellar populations identified in a given dwarf galaxy. Since all stars in our simulated
dwarf galaxies are identical and are initially distributed in the form of an exponential disk, we create such two
populations artificially by dividing the stars in the initial disk simply into two bins with radii $r < r_{\rm sep}$ and
$r > r_{\rm sep}$, where $r_{\rm sep}$ is the radius separating the two populations. We will from now on call these
populations Population 1 (or Pop 1 for short) and Population 2 (Pop 2), respectively. Although artificial, this
simple scheme mimics to some extent what actually happens when new gas is accreted by an isolated dIrr galaxy,
sinks towards the centre and forms a new population of stars there. Since our simulations do not include gas
dynamics and star formation processes, the detailed properties of stellar populations may change with
respect to the ones used here once those are included; we will address these issues in future work.

\begin{figure}
\begin{center}
    \leavevmode
    \epsfxsize=8.2cm
    \epsfbox[60 340 495 780]{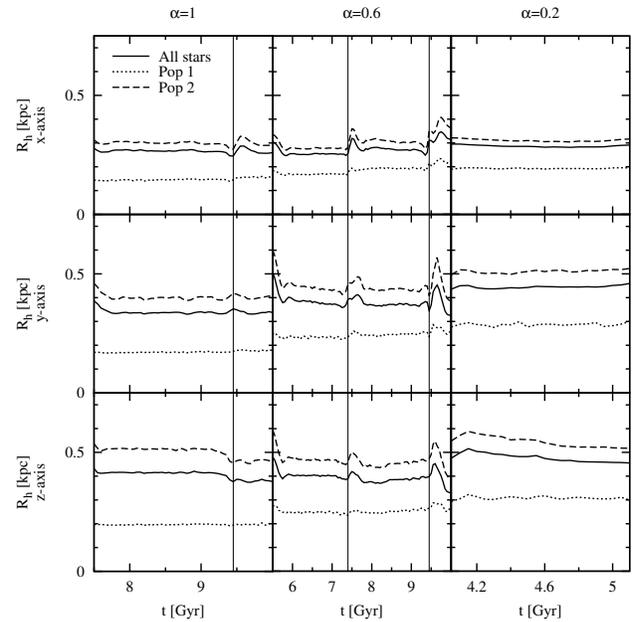}
\end{center}
\caption{The fitted values of the half-light radii for all stars (solid lines) and two populations (dashed and
dotted lines) as a function of time. The three columns from the left to the right show results for
$\alpha = 1, 0.6$ and 0.2, respectively. In rows we present the values fitted for different lines of sight,
along the $x$, $y$ and $z$ axis of the stellar component.
Thin vertical lines indicate pericentre passages.}
\label{populations}
\end{figure}

A most natural choice for $r_{\rm sep}$ would seem to be the
initial 3D half-light radius since then the two populations are equally numerous. Having divided the initial
distribution in two halves in this way we traced the evolution of each one and found that the stars from the
inner Pop 1 soon populate distances $r > r_{\rm sep}$, while stars from the outer Pop 2 migrate inward to
fill the inner gap and form a core-like distribution there. (For a more detailed discussion, including different
orbits, see {\L}okas, Kowalczyk \& Kazantzidis 2012c.) The distributions of the two populations after a few
Gyr of evolution remain
significantly different, but the outer Pop 2 is stripped more strongly and thus is much less numerous,
especially for $\alpha  = 0.2$ where the dwarf is most strongly affected by tides.
To obtain populations of similar size and with profiles well fitted by the Plummer law,
in order to be able to reliably measure the corresponding half-light radii, we
tried smaller values of $r_{\rm sep}$ and finally chose $r_{\rm sep} = 0.2$ kpc.

For each of the two populations selected in this way we calculated its stellar number density profile for a given
line of observation. Examples of such profiles, one for each run with different $\alpha$, are shown in
Figure~\ref{profiles}. We then fitted to such data the projected Plummer law
(\ref{plummer}), exactly as was done previously for the
sample of all stars. The results in terms of the fitted half-light radius for the two populations, $R_{{\rm h},1}$ and
$R_{{\rm h},2}$, are shown as dotted and dashed lines respectively in Figure~\ref{populations} as a function of time
for all outputs selected for the analysis and all lines of sight. As expected, the values of $R_{{\rm h},1}$ for
the more concentrated
Pop 1 are always smaller and $R_{{\rm h},2}$ for the more extended Pop 2 larger than $R_{\rm h}$ for all stars.

\begin{figure}
\begin{center}
    \leavevmode
    \epsfxsize=8.2cm
    \epsfbox[65 340 550 780]{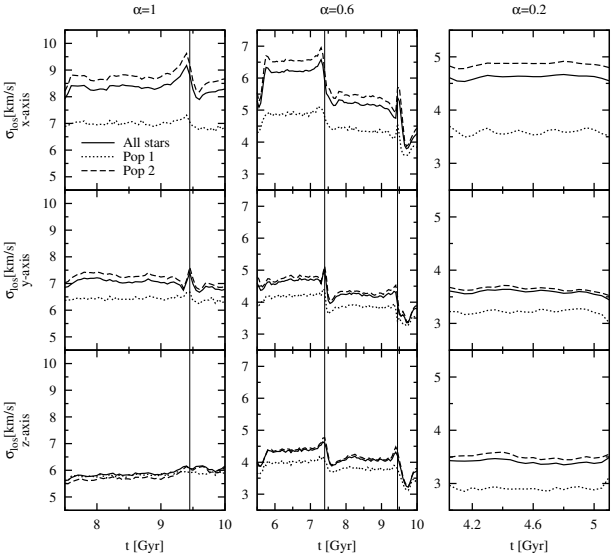}
\end{center}
\caption{The values of the line-of-sight velocity dispersion measured within
half-light radii for all stars (solid lines) and two populations (dashed and
dotted lines) as a function of time. The three columns from the left to the right show results for
$\alpha = 1, 0.6$ and 0.2, respectively. In rows we present the values for different lines of sight,
along the $x$, $y$ and $z$ axis of the stellar component.
Thin vertical lines indicate pericentre passages.}
\label{sigma_populations}
\end{figure}

\begin{figure}
\begin{center}
    \leavevmode
    \epsfxsize=8.2cm
    \epsfbox[65 340 550 780]{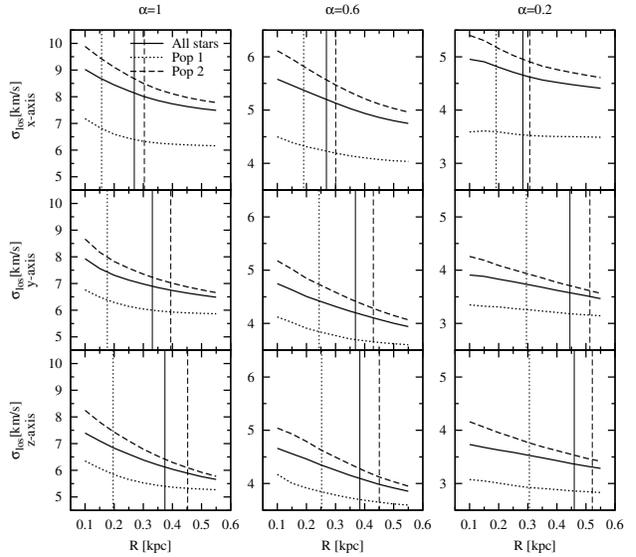}
\end{center}
\caption{Examples of the profiles of the line-of-sight velocity dispersion for all stars (thicker solid lines)
and two populations (thicker dashed and dotted lines). The three columns from the left to the right show results
for a single simulation output at times $t=9.7$, 8.95 and 4.9 Gyr, for $\alpha = 1, 0.6$ and 0.2,
respectively. In rows we present results for different lines of sight, along the $x$, $y$ and $z$ axis of the stellar
component. The thinner vertical lines of the same type indicate the corresponding values of the half-light radii.}
\label{dispersions}
\end{figure}

\begin{figure*}
\begin{center}
    \leavevmode
    \epsfxsize=17cm
    \epsfbox[75 225 560 590]{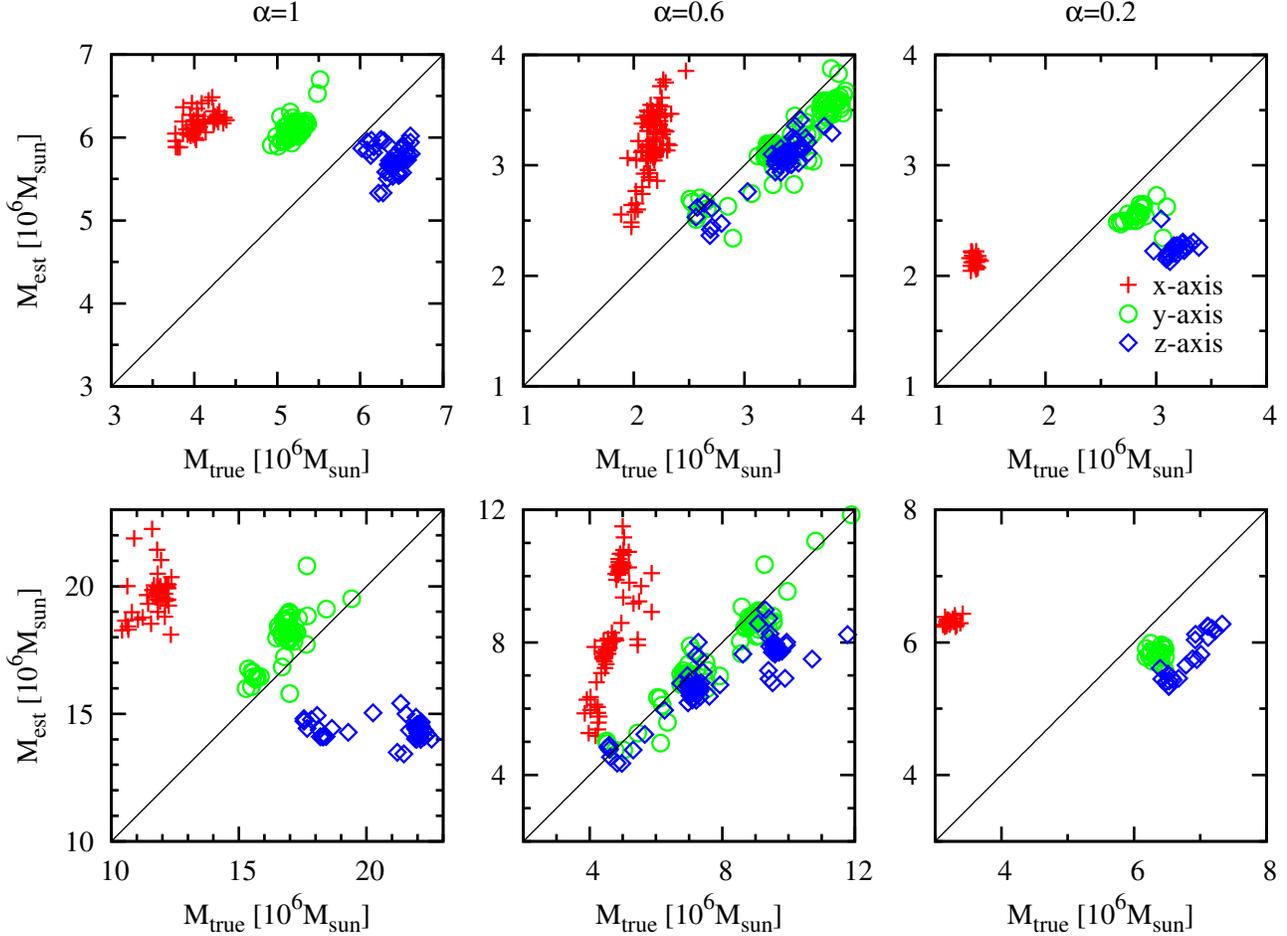}
\end{center}
\caption{The same as Figure~\ref{masses} but using data only for stars in Pop 1 (upper panel) and only
in Pop 2 (lower panel).}
\label{masses-pop}
\end{figure*}

To estimate the masses and slopes of the mass distribution (see the following section)
we also need to measure the line-of-sight velocity
dispersions of the stars belonging to each population within their respective half-light radii $R_{{\rm h},1}$ and
$R_{{\rm h},2}$. These velocity dispersions are plotted as dotted and dashed lines in
Figure~\ref{sigma_populations} as a function of time. Interestingly, the velocity dispersions follow the hierarchy
of half-light radii: the values of $\sigma_{\rm los}$ for the more concentrated
Pop 1 are always smaller and for the more extended Pop 2 larger than the dispersion for all stars for all lines of
sight and all $\alpha$. The only exception occurs for $\alpha=1$ and observation along the shortest $z$ axis of the
stellar component, where the values of $\sigma_{\rm los}$ are very similar for all populations. This can be
understood by referring to Figure~\ref{dispersions} where we show examples of the line-of-sight velocity dispersion
profiles for selected outputs, at times $t=9.7$, 8.95 and 4.9 Gyr, for $\alpha = 1, 0.6$ and 0.2,
respectively (the same outputs were used in Figure~\ref{profiles}) and different lines of sight.

For all cases the velocity dispersion profile decreases with the projected radius $R$ in the region of $R$ we probe,
although more steeply for the
less concentrated populations. In addition, the velocity dispersion profile for Pop 1 is always below the other two,
as expected from the single-value measurements of $\sigma_{\rm los}$ shown in Figure~\ref{sigma_populations}
(an effect also noticed by McConnachie, Pe\~narrubia \& Navarro 2007). Although the profiles show similar behaviour
for all $\alpha$ and lines of sight, the single values used in the mass estimator and shown in
Figure~\ref{sigma_populations} depend also on the actual half-light
radius within which they are measured. In Figure~\ref{dispersions} we indicate the corresponding values of
$R_{\rm h}$ by vertical lines of the same type as used to plot the dispersion profiles.
The hierarchy of half-light radii is a
little different for different $\alpha$ and lines of sight, for example for $\alpha=1$ the differences between the
values of $R_{\rm h}$ for different populations
grow systematically when going from observation along the $x$ to $z$ axis (see the vertical lines in the left-column
plots of Figure~\ref{dispersions}), which results in values of
averaged $\sigma_{\rm los}$ becoming similar for all populations.
This behaviour is a little different for $\alpha=0.6$ and 0.2 because then
the dwarfs are more prolate than triaxial (see also axis ratios in Figure~\ref{transformation}).

In order to verify how working with sub-populations affects the mass estimates, we performed a similar comparison
of estimated versus true masses as was done using all stars in the previous section. First, we calculated the masses
from estimator (\ref{wolf}) for each population separately, using the measured values of half-light radii
$R_{{\rm h},1}$ and $R_{{\rm h},2}$ and the corresponding velocity dispersions.
Then, we measured the actual masses of the simulated dwarfs
within the half-light radius of a given population. The two measurements are compared in
Figure~\ref{masses-pop} where the two rows show results for Pop 1 and Pop 2 respectively. The corresponding values
of $M_{\rm est}/M_{\rm true}$ for the two populations (means and dispersions) are given in the second and third
row for each $\alpha$ in Table~\ref{estimated}. We see that, as in the case of using all stars, masses are always
overestimated when the observation is along the $x$ axis of the stellar component and underestimated if the observation
is along $z$. In particular, for observations along $x$, when using the data for Pop 2 (Pop 1) the mass is more (less)
overestimated than for all stars. In addition, some bias also occurs for the line of sight along the
intermediate axis $y$ for $\alpha=1$.

\begin{figure}
\begin{center}
    \leavevmode
    \epsfxsize=8.2cm
    \epsfbox[70 250 500 780]{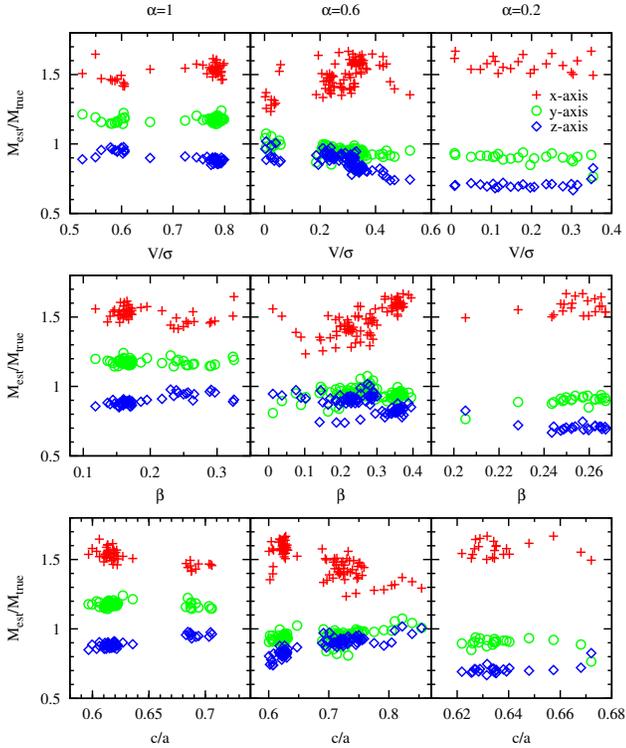}
\end{center}
\caption{The same as Figure~\ref{massratios} but for the more concentrated stellar population Pop 1.}
\label{massratios-pop1}
\end{figure}

\begin{figure}
\begin{center}
    \leavevmode
    \epsfxsize=8.2cm
    \epsfbox[70 250 500 780]{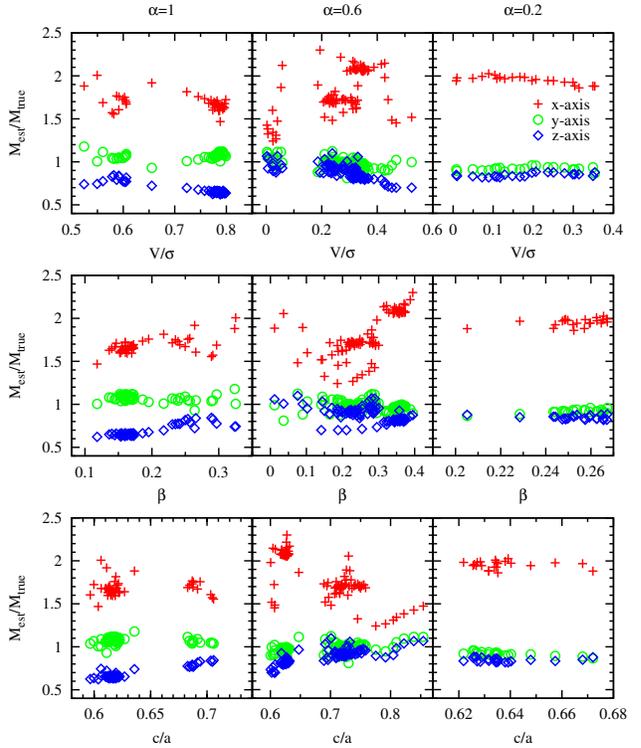}
\end{center}
\caption{The same as Figure~\ref{massratios} but for the less concentrated stellar population Pop 2.}
\label{massratios-pop2}
\end{figure}

In Figures~\ref{massratios-pop1} and \ref{massratios-pop2} we look again at the dependence of $M_{\rm est}/M_{\rm true}$
found for the two populations
on the dwarf properties at a given stage, in terms of $V/\sigma$, $\beta$ and $c/a$. The trends with these parameters
turn out to be similar as for the whole population of stars, namely $M_{\rm est}/M_{\rm true}$ obtained for different
lines of sight converge as dwarfs become more spherical.

\section{Estimating slopes}

In this section we finally measure the slopes of the mass distribution of our simulated dwarfs using the estimator
proposed by WP11
\begin{equation}         \label{wp}
	\Gamma_{\rm est} = \frac{\Delta \log M}{\Delta \log r}
	= 1 + \frac{\log (\sigma^2_{\rm los,1}/\sigma^2_{\rm los,2})}{\log (R_{\rm h,1}/R_{\rm h,2})}
\end{equation}
where $\sigma_{{\rm los},i}$ is the line-of-sight velocity dispersion measured within $R_{{\rm h},i}$ for the
$i$-th population. The estimated slopes calculated in this way were then compared to those measured
directly from the simulation data. This true value $\Gamma_{\rm true}$ is obtained by taking in $\Delta \log M$ in
equation (\ref{wp}) the true masses within $r_{\rm h,1} = f R_{\rm h,1}$ and $r_{\rm h,1} = f R_{\rm h,1}$.
This turns out to be equivalent to a high accuracy to the direct fitting of the mass slope of the simulated dwarfs
at a radius equally distant from $r_{\rm h,1}$ and $r_{\rm h,2}$ in log $r$, that is at
$\log r = (\log r_{\rm h,1} + \log r_{\rm h,2})/2$.

\begin{figure*}
\begin{center}
    \leavevmode
    \epsfxsize=17cm
    \epsfbox[70 410 500 590]{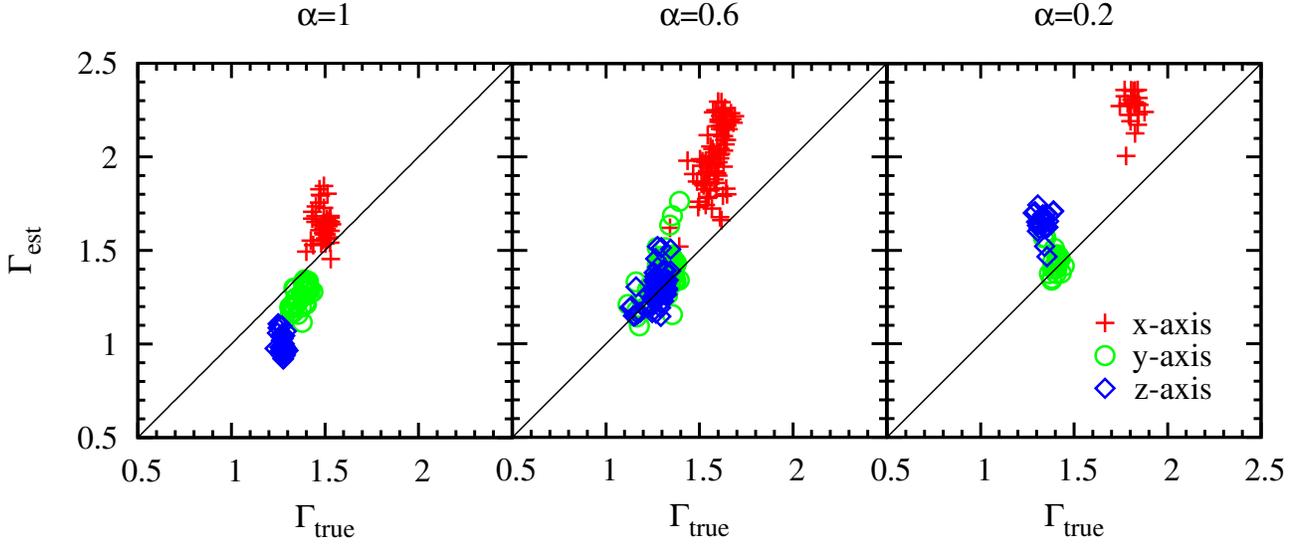}
\end{center}
\caption{Comparison between the mass slopes $\Gamma$ obtained using estimator (\ref{wp})
and by direct measurement of masses at the two half-light radii in the simulation.
The three panels from the left to the right show results for $\alpha = 1, 0.6$ and 0.2,
respectively, and the diagonal line in each panel corresponds to the equality of the $\Gamma$ values.
Symbols of different shape and colour indicate measurements derived from observations performed along different
lines of sight: $x$ (longest axis of the stellar component), $y$ (intermediate axis) and $z$ (shortest axis).}
\label{gammas}
\end{figure*}

The estimated and true values of $\Gamma$ are compared in Figure~\ref{gammas}. Interestingly, in spite of the
large scatter in mass estimates obtained from the kinematic data for the two populations, the estimated $\Gamma$
values for a given $\alpha$ and line of sight are confined to a very tight region. This
demonstrates that the scatter in masses partially cancels out in the estimator (\ref{wp}) and thus speaks in
favour of the method. However, $\Gamma_{\rm est}$ can be overestimated as well as underestimated depending on the
line of sight and in particular it is {\em always\/} overestimated if the line of sight is along the longest
axis of the stellar distribution. The exact values of the means and dispersions of
$\Gamma_{\rm est}/\Gamma_{\rm true}$ are given in the fourth row for each $\alpha$ in Table~\ref{estimated}.

Let us note that the agreement between $\Gamma_{\rm est}$ and $\Gamma_{\rm true}$ is best and almost perfect
for the dwarfs observed along the intermediate $y$ axis of the stellar component, similarly as in the case of
mass estimates.
Interestingly, the behaviour of $\Gamma_{\rm est}/\Gamma_{\rm true}$ for observations along the $z$ axis is
different for different $\alpha$. In particular, the ratio is below unity for $\alpha=1$, of the order of unity
for $\alpha=0.6$ and above unity $\alpha=0.2$. This anomalous behaviour of the $\alpha=0.2$ case
can be traced to a slightly bigger difference between
$\sigma^2_{\rm los,1}$ and $\sigma^2_{\rm los,2}$ for observation along $z$ compared to $y$ axis
(see Figure~\ref{sigma_populations}). While the
differences between $R_{\rm h,1}$ and $R_{\rm h,2}$ are similar for the two lines of sight, and thus
$\Gamma_{\rm true}$ values are similar, the difference in velocity dispersion of the two populations results in
a larger $\Gamma_{\rm est}$ for the observation along $z$.

\begin{figure}
\begin{center}
    \leavevmode
    \epsfxsize=8.1cm
    \epsfbox[70 250 500 780]{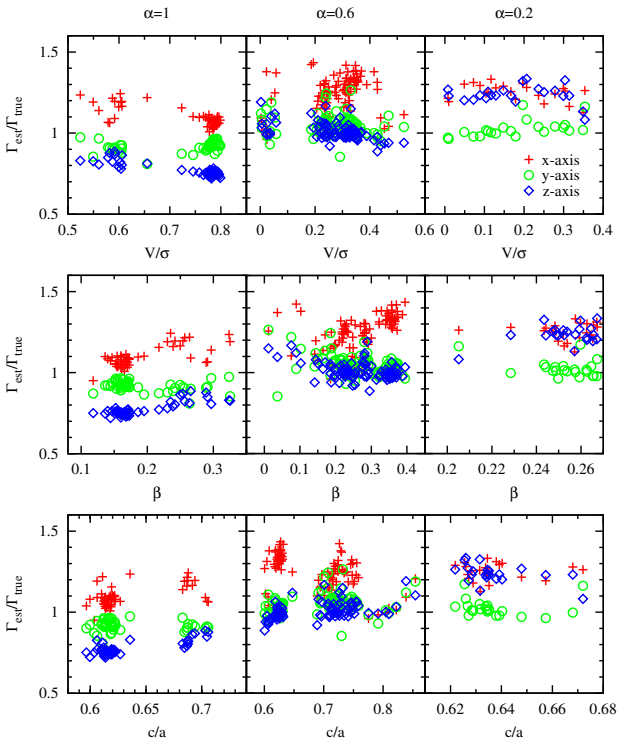}
\end{center}
\caption{The ratio of the estimated to true slope $\Gamma$ as a function of the amount of rotation $V/\sigma$
(upper row), the anisotropy parameter $\beta$ (middle row) and shape in terms of the ratio of the shortest
to longest axis of the stellar component $c/a$ (lower row). The columns from the left to the right show results
for $\alpha = 1, 0.6$ and 0.2, respectively. Symbols of different shape and colour indicate observations performed
along different lines of sight: $x$ (longest axis of the stellar component), $y$ (intermediate axis) and $z$
(shortest axis).}
\label{gammaratios}
\end{figure}

Before we conclude, let us again look at the dependence of the ratio $\Gamma_{\rm est}/\Gamma_{\rm true}$ on
the properties of the dwarf: the amount of rotation, anisotropy and shape (Figure~\ref{gammaratios}).
The trends turn out to be similar as
in the case of mass estimates for the whole population: $\Gamma_{\rm est}/\Gamma_{\rm true}$ tend to unity
for more spherical stellar components.

\section{Discussion}

Using collisionless $N$-body simulations of dwarf galaxies orbiting the Milky Way we created numerical models
of dwarfs with non-spherical stellar components and measured their properties as is usually done in observations
of real dSph galaxies in the Local Group. The use of high-resolution simulations allowed us to construct
kinematic samples almost free of the usual statistical errors associated with observational uncertainties.
The stars in the simulated dwarfs have positions and velocities known to arbitrary accuracy and the membership
of the stars in a given population is known by definition. The statistical errors on the inferred values of
the half-light radii and velocity dispersions, such as sampling errors, are very small due to a large number
of stars available. These facts allowed us to study the systematic errors involved in the determination of the
masses and density slopes of the dwarfs.

We have demonstrated that the inferred values of the mass contained within a radius of the order of a half-light radius
of the stars can be over- or underestimated depending on the line of sight along which the observation is performed.
In particular, the masses are always larger than the real ones by up to a factor of two if the dwarfs are observed
along the longest axis of the stellar component. When observed along the shortest axis, the masses are underestimated
by at most 30 percent. The bias is weakest for the observation along the intermediate axis and in this case the masses
are recovered with a very good accuracy.

The most important conclusion of this work is related to the inferences concerning the slope of mass distribution in
dSph galaxies, an issue which has attracted a lot of attention among the modellers in recent years. Using the same
mock data samples as were used to study the accuracy of mass estimates we studied the reliability of the method
recently proposed by WP11 to infer the slope of mass distribution $\Gamma$. Our main result is summarized in
Figure~\ref{gammas}. We demonstrated that the slope of mass distribution estimated using this method
can be under- and overestimated depending on the line of sight along which the observation is performed.
This has immediate consequences for the inferences concerning the slope of the density profile in dSph galaxies.

Let us define the limiting inner slope of the density profile as $\gamma (r \rightarrow 0)=\gamma_0$. The analogous
value of the mass slope will be $\Gamma (r \rightarrow 0)=\Gamma_0$. As discussed by WP11, in the limit of
small radii
\begin{equation}	\label{gammarelation}
	\gamma_0 = 3 - \Gamma_0 < 3 - \Gamma
\end{equation}
because for any radius $r>0$ we have $\Gamma_0 > \Gamma(r)$. For example, taking $\Gamma_{\rm  est}=2.3$ found for
$\alpha=0.6$ and observation along the $x$ axis (see the middle panel Figure~\ref{gammas}) leads to the constraint
$\gamma_0<0.7$
while taking the corresponding true value $\Gamma_{\rm  true}=1.6$ gives $\gamma_0<1.4$, thus a result entirely
consistent with the presence of an inner cusp. Therefore using an overestimated value of $\Gamma_{\rm  est}$ in
equation (\ref{gammarelation}) may lead us to infer a presence of a core-like profile when no such profile is
really there.

\begin{figure}
\begin{center}
    \leavevmode
    \epsfxsize=8.3cm
    \epsfbox[70 329 515 674]{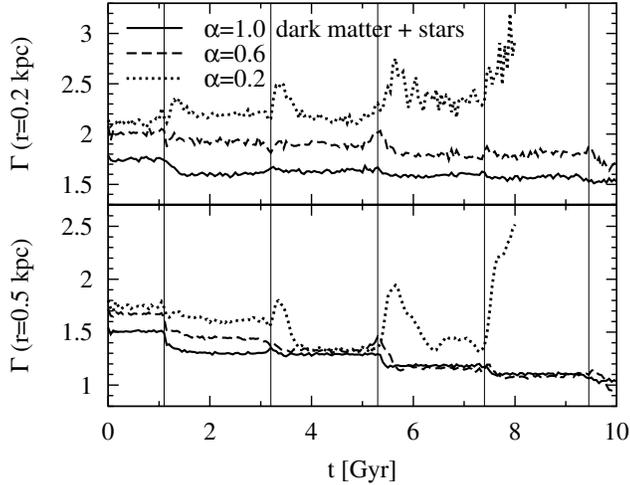}
\end{center}
\caption{The evolution of the slope of the total (dark matter + stars)
mass profile in time measured at the radius $r=0.2$ (upper
panel) and $r=0.5$ kpc (lower panel) from the centre of the dwarf.
The solid, dashed and dotted lines show slopes measured for dwarfs with
haloes of different initial inner slope $\alpha = 1, 0.6$ and 0.2, respectively. Thin vertical lines indicate
pericentre passages.}
\label{Gamma}
\end{figure}

We further illustrate this conclusion by showing in the two lower panels of Figure~\ref{evolution} and in
Figure~\ref{Gamma} the actual slopes
of the total (dark matter and stars) density and mass profiles, respectively. The slopes are shown as a function
of time for radii bracketing our range of half-light radii (0.2 and 0.5 kpc) for $\alpha=0.2$, 0.6 and 1. As
we can see from Figure~\ref{evolution}, at $r=0.2$ even for $\alpha=0.2$ the total density slope $-\gamma<-1$
at most times until the third pericentre and thus indeed no core is present at this scale. Note that $r=0.2$ is
of the order of 3-4 softening scales for dark matter particles in our simulations and thus is the smallest scale
where we can reliably determine the total density profile. The method of WP11 obviously is supposed to
give the slope of the total mass profile and the inferences concerning the most interesting question of the slope of
the dark matter distribution can only be made under the assumption that dark matter dominates strongly
over stars everywhere in the dwarf galaxy. This may not always be the case, especially in the central part of dwarfs.
For our simulated dwarfs the slopes do differ as confirmed by comparison between the first and third row panels of
Figure~\ref{evolution}.

Figure~\ref{Gamma} can also be viewed as a consistency check for the
$\Gamma_{\rm true}$ values shown in Figure~\ref{gammas}. The values of $\Gamma_{\rm true}$ should fall between
the values shown in the two panels of Figure~\ref{Gamma} since $\Gamma_{\rm true}$ are measured at radii from
the range bracketed by half-light radii of the two stellar populations and these are typically between 0.2 and 0.5 kpc
(see Figure~\ref{populations}). Indeed, for the selected outputs we used the values of $\Gamma_{\rm true}$ from
Figure~\ref{gammas} are in the range $1.2-1.5$, $1.1-1.7$ and $1.3-1.8$ respectively for $\alpha=1$, 0.6 and 0.2,
in agreement with values plotted in Figure~\ref{Gamma}.

WP11 applied their method to the data for the Fornax and Sculptor dSph galaxies obtaining $\Gamma$ values of
2.61 and 2.95 with a rather small error of the order of 0.4.
Thus, according to equation (\ref{gammarelation}), they inferred the presence of cores in these galaxies:
$\gamma_0 < 0.39$ and $\gamma_0 < 0.05$. Given our results, these inferences can be systematically biased
towards the presence of the core if the dwarfs are observed along the major axis of the stellar component.
Our results thus weaken the tension between the findings of WP11 and recent results of simulations
(Governato et al. 2012)
where such extended cores tend to form only in significantly more massive galaxies.

\begin{table}
\caption{Estimated masses and slopes versus true values. }
\label{estimated}
\begin{center}
\begin{tabular}{lccc}
Quantity & along $x$ & along $y$ & along $z$      \\
\hline
						    &  & $\alpha=1$ &   \\
$M_{\rm est}/M_{\rm true} (f R_{\rm h})$            & $1.61 \pm 0.08$ & $1.05 \pm 0.03$ & $0.72 \pm 0.06$ \\
$M_{\rm est}/M_{\rm true} (f R_{\rm h,1})$          & $1.52 \pm 0.05$ & $1.18 \pm 0.02$ & $0.90 \pm 0.03$ \\
$M_{\rm est}/M_{\rm true} (f R_{\rm h,2})$          & $1.68 \pm 0.09$ & $1.07 \pm 0.04$ & $0.69 \pm 0.07$ \\
$\Gamma_{\rm est}/\Gamma_{\rm true} (f R_{\rm h,i})$  & $1.10 \pm 0.06$ & $0.92 \pm 0.03$ & $0.77 \pm 0.04$ \\
\\
						    &  & $\alpha=0.6$ & \\
$M_{\rm est}/M_{\rm true} (f R_{\rm h})$            & $1.72 \pm 0.22$ & $0.96 \pm 0.05$ & $0.87 \pm 0.07$ \\
$M_{\rm est}/M_{\rm true} (f R_{\rm h,1})$          & $1.48 \pm 0.11$ & $0.95 \pm 0.04$ & $0.88 \pm 0.06$ \\
$M_{\rm est}/M_{\rm true} (f R_{\rm h,2})$          & $1.81 \pm 0.25$ & $0.98 \pm 0.06$ & $0.88 \pm 0.08$ \\
$\Gamma_{\rm est}/\Gamma_{\rm true} (f R_{\rm h,i})$  & $1.25 \pm 0.11$ & $1.05 \pm 0.06$ & $1.01 \pm 0.05$ \\
\\
						    &  & $\alpha=0.2$ & \\
$M_{\rm est}/M_{\rm true} (f R_{\rm h})$            & $1.86 \pm 0.04$ & $0.91 \pm 0.02$ & $0.81 \pm 0.02$ \\
$M_{\rm est}/M_{\rm true} (f R_{\rm h,1})$          & $1.57 \pm 0.06$ & $0.90 \pm 0.04$ & $0.71 \pm 0.03$ \\
$M_{\rm est}/M_{\rm true} (f R_{\rm h,2})$          & $1.95 \pm 0.04$ & $0.92 \pm 0.02$ & $0.85 \pm 0.02$ \\
$\Gamma_{\rm est}/\Gamma_{\rm true} (f R_{\rm h,i})$  & $1.25 \pm 0.05$ & $1.03 \pm 0.05$ & $1.24 \pm 0.06$ \\
\hline
\end{tabular}
\end{center}
\end{table}

Our results suggest that the method could be more reliable and the systematics of the uncertainties could be more under
control if we knew the orientation of the studied object with respect to our line of sight. Although most of dSph
galaxies are known to be non-spherical from their surface density maps, their orientation remains unknown. If most
of the dSphs indeed formed via tidal stirring, they should remain non-spherical until the present except for those
that evolved on tight enough orbits for a long enough time (Kazantzidis et al. 2011). The orientation of the major
axis of the stellar component should then be random, as the dwarfs are supposed to tumble with a period much shorter
than the orbital period of their motion around the Milky Way (see figure 4 in {\L}okas et al. 2011).

An interesting exception where the orientation of the dwarf could be determined is
the case of the Sagittarius dwarf. According to the model proposed by {\L}okas et al.
(2010b) the dwarf is now close to the second pericentre of its orbit around the Milky Way and forms a bar oriented
perpendicular to our line of sight. The Sagittarius dwarf could thus become a promising target for the unbiased
application of the method but the case requires further studies to confirm its orientation and identify multiple
stellar populations among its stars.

Despite the biases inherent in this method when applied to a single dSph, which is necessarily observed along one
line of sight, it can still be useful when applied to a larger sample of nearby dSph galaxies which should be
oriented randomly with respect to our position in the Galaxy. The biases are then expected to diminish and
a mean result can be closer to the truth than for a single object. This averaging may work if the dwarfs were all
formed in a similar fashion and possess similar properties, in particular in terms of the dark matter distribution.
Given their different masses, luminosities and star formation histories this may not however be the case.

\section*{Acknowledgements}

This research was partially supported by the Polish National Science
Centre under grant NN203580940. KK acknowledges the summer student program of the Copernicus Center
in Warsaw. The numerical simulations used in this work were performed at the
Ohio Supercomputer Center (http://www.osc.edu).


\begin{thebibliography}{}

\bibitem[{Breddels et al.}(2012)]{breddels} Breddels M. A., Helmi A., van den Bosch R. C. E., van de Ven G.,
	Battaglia G., 2012, submitted to MNRAS, arXiv:1205.4712
\bibitem[{Bruzual \& Charlot}(2003)]{bc03} Bruzual G., Charlot S., 2003, MNRAS, 344, 1000
\bibitem[{Chaname et al.}(2008)]{chan} Chanam\'e J., Kleyna J., van der Marel R., 2008, ApJ, 682, 841
\bibitem[{Cole et al.}(2012)]{cole} Cole D. R., Dehnen W., Read J. I., Wilkinson M. I.,	2012, MNRAS, 426, 601
\bibitem[{de Blok et al.}(2001)]{bmr} de Blok W. J. G., McGaugh S. S., Rubin V. C., 2001, AJ, 122, 2396
\bibitem[{Diemand et al.}(2007)]{dkm} Diemand J., Kuhlen M., Madau P., 2007, ApJ, 667, 859
\bibitem[{Goerdt et al.}(2006)]{goerdt} Goerdt T., Moore B., Read J. I., Stadel J., Zemp M., 2006, MNRAS, 368, 1073
\bibitem[{Governato et al.}(2010)]{gov10} Governato F. et al., 2010, Nature, 463, 203
\bibitem[{Governato et al.}(2012)]{gov12} Governato F. et al., 2012, MNRAS, 422, 1231
\bibitem[{Jardel \& Gebhardt}(2012)]{jg12} Jardel J. R., Gebhardt K., 2012, ApJ, 746, 89
\bibitem[{Kazantzidis et al.}(2004)]{kaz04} Kazantzidis S., Mayer L., Mastropietro C., Diemand J.,
	Stadel J., Moore B., 2004, ApJ, 608, 663
\bibitem[{Kazantzidis et al.}(2011)]{kaz11} Kazantzidis S., {\L}okas E. L., Callegari S., Mayer L.,
	Moustakas L. A., 2011, ApJ, 726, 98
\bibitem[{Kazantzidis et al.}(2013)]{kaz13} Kazantzidis S., {\L}okas E. L., Mayer L., 2013, ApJ, 764, L29
\bibitem[{Klimentowski et al.}(2007)]{k07} Klimentowski J., {\L}okas E. L., Kazantzidis S., Prada F.,
	Mayer L., Mamon G. A., 2007, MNRAS, 378, 353
\bibitem[{Klimentowski et al.}(2009a)]{k09a} Klimentowski J., {\L}okas E. L., Kazantzidis S.,
	Mayer L., Mamon G. A., 2009a, MNRAS, 397, 2015
\bibitem[{Klimentowski et al.}(2009b)]{k09b} Klimentowski J., {\L}okas E. L., Kazantzidis S.,
	Mayer L., Mamon G. A., Prada F., 2009b, MNRAS, 400, 2162
\bibitem[{Klimentowski et al.}(2010)]{k10} Klimentowski J., {\L}okas E. L., Knebe A., Gottl\"ober S., Martinez-Vaquero
	L. A., Yepes G., Hoffman Y., 2010, MNRAS, 402, 1899
\bibitem[{Lokas}(2002)]{lo02} {\L}okas E. L., 2002, MNRAS, 333, 697
\bibitem[{Lokas \& Mamon}(2003)]{lm03} {\L}okas E. L., Mamon G. A., 2003, MNRAS, 343, 401
\bibitem[{Lokas et al.}(2005)]{lmp} {\L}okas E. L., Mamon G. A., Prada F., 2005, MNRAS, 363, 918
\bibitem[{Lokas et al.}(2010a)]{lo10a} {\L}okas, E. L., Kazantzidis, S., Klimentowski, J., Mayer, L., \& Callegari, S.
        2010a, ApJ, 708, 1032
\bibitem[{Lokas et al.}(2010b)]{lo10b} {\L}okas, E. L., Kazantzidis, S., Majewski, S. R., Law, D. R., Mayer, L.,
	\& Frinchaboy, P. M. 2010b, ApJ, 725, 1516
\bibitem[{Lokas et al.}(2011)]{lokas11} {\L}okas E. L., Kazantzidis S., Mayer L., 2011, ApJ, 739, 46
\bibitem[{Lokas et al.}(2012a)]{lokas12a} {\L}okas E. L., Majewski S. R., Kazantzidis S., Mayer L., Carlin J. L.,
	Nidever D. L., Moustakas L. A., 2012a, ApJ, 751, 61
\bibitem[{Lokas et al.}(2012b)]{lokas12b} {\L}okas E. L., Kazantzidis S., Mayer L., 2012b, ApJ, 751, L15
\bibitem[{Lokas et al.}(2012c)]{lokas12c} {\L}okas E. L., Kowalczyk K., Kazantzidis S., 2012c,
	in Bono G., de Grijs R., Omizzolo A., eds, Proc. EWASS 2012 Symp. 6, Stellar Populations 55 years
	after the Vatican Conference. Memorie della Societa Astronomica Italiana, in press, arXiv:1212.0682
\bibitem[{Mateo}(1998)]{mat98} Mateo M. L., 1998, ARA\&A, 36, 435
\bibitem[{McConnachie et al.}(2007)]{mpn07} McConnachie A. W., Pe\~narrubia J., Navarro J. F., 2007, MNRAS, 380, L75
\bibitem[{Mayer et al.}(2001)]{may01} Mayer L., Governato F., Colpi M., Moore B., Quinn T., Wadsley J.,
	Stadel J., Lake G., 2001, ApJ, 559, 754
\bibitem[{Mayer et al.}(2007)]{may07} Mayer L., Kazantzidis S., Mastropietro C., Wadsley J., 2007, Nature, 445, 738
\bibitem[{Mo et al.}(1998)]{mo98} Mo  H. J., Mao S., White S. D. M., 1998, MNRAS, 295, 319
\bibitem[{Navarro et al.}(1997)]{nfw97} Navarro J. F., Frenk C. S., White S. D. M., 1997, ApJ, 490, 493 (NFW)
\bibitem[{Oh et al.}(2011)]{oh11} Oh S.-H. et al., 2011, AJ, 142, 24
\bibitem[{Sanchez-Conde et al.}(2007)]{sc07} S\'anchez-Conde M. A., Prada F., {\L}okas E. L., G\'omez M. E.,
	Wojtak R., Moles M., 2007, Phys. Rev. D, 76, id. 123509
\bibitem[{Spergel \& Steinhardt}(2000)]{ss00} Spergel D. N., Steinhardt P. J., 2000, Phys. Rev. Letters, 84, 3760
\bibitem[{Stadel}(2001)]{sta01} Stadel J. G., 2001, PhD thesis, Univ. of Washington
\bibitem[{Vogelsberger et al.}(2012)]{vzl} Vogelsberger M., Zavala J., Loeb A., 2012, MNRAS, 423, 3740
\bibitem[{Walker \& Penarrubia}(2011)]{wp11} Walker M. G., Pe\~narrubia J., 2011, ApJ, 742, 20
\bibitem[{Widrow \& Dubinski}(2005)]{wd05} Widrow L. M., Dubinski J., 2005, ApJ, 631, 838
\bibitem[{Wolf et al.}(2010)]{wolf} Wolf J., Martinez G. D., Bullock J. S., Kaplinghat M., Geha M., Mu\~noz R. R.,
	Simon J. D., Avedo F. F., 2010, MNRAS, 406, 1220

\end{thebibliography}
\end{document}